\documentclass[twocolumn,showpacs,preprintnumbers,amsmath,amssymb]{revtex4}

\usepackage{graphicx}
\usepackage{dcolumn}
\usepackage{bm}
\usepackage{color}

\begin{document}

\preprint{APS/123-QED}

\title{Unconventional critical scaling of magnetization \\in uranium ferromagnetic superconductors UGe$_2$ and URhGe\footnote{Phys. Rev. B {\bf 89}, 064420 (2014).}}

\author{Naoyuki Tateiwa$^{1}$}
\email{tateiwa.naoyuki@jaea.go.jp} 
\author{Yoshinori Haga$^{1}$}%
\author{Tatsuma D. Matsuda$^{1,2}$}%
\author{Etsuji Yamamoto$^{1}$}%
\author{Zachary Fisk$^{1,3}$}

\affiliation{
$^{1}$Advanced Science Research Center, Japan Atomic Energy Agency, Tokai, Naka, Ibaraki 319-1195, Japan\\
$^2$Department of Physics, Tokyo Metropolitan University, Hachioji, Tokyo 192-0397, Japan\\
$^{3}$University of California, Irvine, California 92697, USA\\
}
\date{January 22, 2014}

\begin{abstract}
We report a dc magnetization study of the critical phenomenon around the ferromagnetic transition temperature $T_{\rm C}$ in high-quality single crystals of uranium ferromagnetic superconductors UGe$_2$ and URhGe. The critical exponents, $\beta$ for the temperature dependence of the magnetization below $T_{\rm C}$, $\gamma$ for the magnetic susceptibility, and $\delta$ for the magnetic isothermal at $T_{\rm C}$ have been determined with a modified Arrott plot, a Kouvel-Fisher plot, and the scaling analysis. Magnetization in the ferromagnetic state has strong uniaxial magnetic anisotropy in the two compounds. However, the universality class of the critical phenomena do not belong to the three dimensional (3D) Ising system. Although the values of $\beta$ in UGe$_2$ and URhGe are close to those in the 3D magnets, the values of $\gamma$ are close to unity, that expected from the mean field theory. Similar critical exponents have been reported previously for the 3D Ising ferromagnet UIr where superconductivity appears under high pressure. The critical behavior may be limited to a very narrow Ginzburg critical region of ${\Delta}{T_{\rm G}}{\,}{\sim}$ 1 mK because of the strong itinerant character of the $5f$ electrons in the ferromagnetic superconductor UCoGe where the mean field behavior of the magnetization has been reported. The unconventional critical scaling of magnetization in UGe$_2$, URhGe and UIr cannot be explained via previous approaches to critical phenomena. The ferromagnetic correlation between the $5f$ electrons differs from that in the 3D Ising system and this difference may be a key point for the understanding of the ferromagnetism where superconductivity emerges. 

\end{abstract}


\maketitle

\section{Introduction}

 The coexistence of superconductivity and ferromagnetism, considered as a theoretical possibility over 50 years ago by Ginzburg\cite{ginzburg}, has been found in the uranium compounds UGe$_2$\cite{saxena,huxley0}, URhGe\cite{aoki}, and UCoGe\cite{huy}. Extensive theoretical and experimental studies have been carried out\cite{aoki2}. Since the middle of the 1970s, the coexistence has been found in the $4f$-localized systems such as ErRh$_4$B$_4$\cite{moncton,fertig}, Chevrel compound HoMo$_6$S$_8$\cite{ishikawa}, and boron carbide superconductor ErNi$_2$B$_2$C\cite{canfield}. The ferromagnetism and superconductivity of the systems are carried by different electrons: $f$ and $d$ electrons respectively, and the states compete each other. A characteristic feature in the uranium ferromagnetic superconductors is that the same $5f$ electrons of the uranium atoms are responsible for both long-range ordered states. Interesting physical phenomena such as anomalous enhancement of the upper critical field $H_{c2}$ for the superconducting state under high magnetic field may originate from cooperative interplay between the two phases\cite{aoki2}.

 Critical ferromagnetic fluctuations have been thought to induce unconventional superconductivity in the vicinity of a quantum phase transition\cite{fay}. Spin fluctuation theories reveal the importance of the dimensionality of the spin-fluctuation spectrum for the unconventional superconductivity\cite{monthoux1,monthoux2,wang}. In particular, longitudinal spin fluctuations play an important role for spin-triplet superconductivity as was experimentally shown in a recent NMR experiment on UCoGe\cite{hattori}. Indeed, the ferromagnetic phase has strong Ising-type anisotropy in the uranium ferromagnetic superconductors\cite{onuki,aoki2,huy}. Although many studies have been done on the superconductivity and its related phenomena, a systematic and complete description of the ferromagnetic criticality has not been made for the uranium ferromagnetic superconductors. 
 
  Ferromagnetism in the uranium ferromagnetic superconductors is carried by the mobile $5f$ electrons\cite{fujimori}. Itinerant ferromagnetism of the $5f$ electrons may have magnetic properties differing from those in intermetallic compounds of $3d$ transition metals. The relaxation rate for the magnetization density $\Gamma$ in UGe$_2$ and UCoGe does not exhibit the linear Landau damping (${\Gamma}{\,}{\propto}{\,}{{\mbox{\boldmath $q$}}}$) characteristic of the itinerant ferromagnetism described by self-consistent renormalized spin fluctuation (SCR) theory \cite{huxley1,stock,moriya}. Dual nature of the $5f$ electrons between itinerant and localized characters has been suggested in Muon spin rotation spectroscopy and macroscopic experiments in UGe$_2$\cite{yaouanc,sakarya,troc}.  It is necessary then to investigate particular features in the itinerant ferromagnetism of the $5f$ electrons.

  In this paper, we present detailed dc magnetization studies of UGe$_2$ and URhGe to investigate the classical critical phenomenon associated with the ferromagnetic transition. UGe$_2$ shows a second order phase transition from the paramagnetic to the ferromagnetic (FM1) phases at $T_{\rm C}$ = 52.6 K and URhGe orders ferromagnetically at $T_{\rm C}$ = 9.5 K\cite{saxena,huxley0,aoki,aoki2}. Superconductivity appears in the high pressure FM1 phase in UGe$_2$. URhGe has a superconducting transition with transition temperature $T_{sc}$ = 0.2 K at ambient pressure. We study the critical phenomena of the ferromagnetic states in the two compounds where the superconductivity appears at low temperatures. It is found that the universality class for the ferromagnetic transitions in UGe$_2$ and URhGe does not belong to the three dimensional (3D) Ising class expected from the strong uniaxial anisotropy in the magnetization. We find a unique scaling relation which may be inherent to the uranium ferromagnetic superconductors.

 In the vicinity of a second-order magnetic phase transition with Curie temperature $T_{\rm C}$, the divergence of correlation length $\xi$ = ${\xi}_0$ $|1-T/{T_{\rm C}}|^{-{\nu}}$ leads to universal scaling laws for spontaneous magnetization $M_S$ and initial susceptibility ${\chi}$. ${{\nu}}$ is the critical exponent. The mathematic definitions of exponents from magnetization can be described as follows\cite{privman}.   
      \begin{eqnarray}
   {{\chi}}(T){^{-1}}&{\propto}&  {|t|}^{-{{\gamma}}'} {\;}{\;}{\;}{\;}(T<T{_{\rm C}}), {\;}{\;} {|t|}^{-{{\gamma}}} {\;}{\;}(T{_{\rm C}}< T)\\ 
 M{_S}(T) &{\propto}& |t|{^{\beta}}  {\;}  {\;} {\;}{\;}{\;} {\;}(T < T{_{\rm C}})\\ 
  {{\mu}_0}H& {\propto} & {M{_S}}^{1/{\delta}}   {\;} {\;} {\;} (T = T{_{\rm C}})
   \end{eqnarray}

Here, $t$ denotes the reduced temperature $t$ = $1-{T}/{T_{\rm C}}$. Parameters $\beta$, $\gamma$, ${\gamma}'$  and $\delta$ are the critical exponents. Table I shows the theoretical critical exponents for various models. 

     \begin{table}[t]
\caption{\label{tab:table1}%
Critical exponents $\beta$, $\gamma$, $\delta$ for different universality classes\cite{privman}. }
\begin{ruledtabular}
\begin{tabular}{ccccc}
\textrm{}&
\textrm{$\beta$}&
\textrm{$\gamma$}&
\textrm{$\delta$}&\\
\colrule
Mean field &0.5 &1& 3 \\
3D Heisenberg & 0.367 &1.388 &4.78 \\
3D XY & 0.345  &1.316 &4.81 \\
3D Ising & 0.326 &1.238 & 4.80  \\
2D Ising & 0.125 &1.75& 15  \\
\end{tabular}
\end{ruledtabular}
 \end{table}
\section{EXPERIMENT}
High-quality single crystal samples of UGe$_2$ and URhGe have been grown by Czochralski pulling in a tetra arc furnace\cite{onuki,yamamoto}. Magnetization was measured in a commercial superconducting quantum interference (SQUID) magnetometer (MPMS, Quantum Design). The internal magnetic field ${{\mu}_0}H$ was determined by subtracting the demagnetization field $DM$ from the applied magnetic field ${{\mu}_0}H_{ext}$: ${{\mu}_0}H$ = ${{\mu}_0}H_{ext}$ - $DM$. The demagnetizing factor $D$ was calculated from the macroscopic dimensions of the sample. The magnetic field was applied along the magnetic easy $a$ and $c$-axes of the orthorhombic structure of UGe$_2$ and URhGe, respectively. We have determined the critical exponents in the compounds using a modified Arrott plot, critical isotherm analysis, a Kouvel-Fisher plot, and scaling analysis. 

\section{RESULTS}

\subsection{Modified Arrott plot and critical isotherm}
    \begin{figure}[b]
\includegraphics[width=8.5cm]{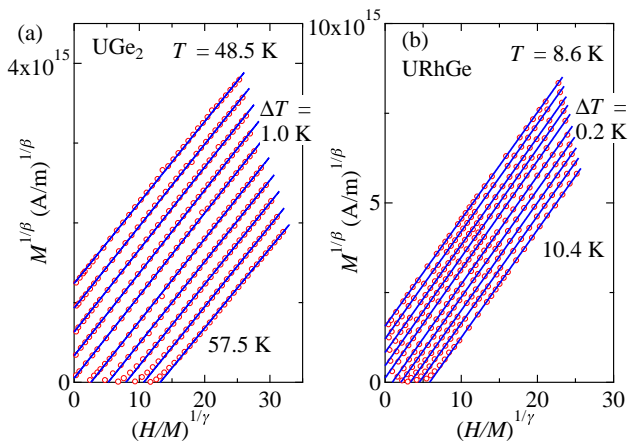}
\caption{\label{fig:epsart}(Color online) Modified Arrott plot of magnetization in (a)UGe$_2$ for 48.5 K $\le$ $T$ $\le$ 57.5 K and 0.1 T $\le$ ${{\mu}_0}H$ $\le$ 6.0 T and in (b) URhGe for 8.6 K $\le$ $T$ $\le$ 10.4 K and 0.1 T $\le$ ${{\mu}_0}H$ $\le$ 2.0 T. Blues lines show fits to the data with the equation (4).}
\end{figure} 
    \begin{figure}[]
\includegraphics[width=8.5cm]{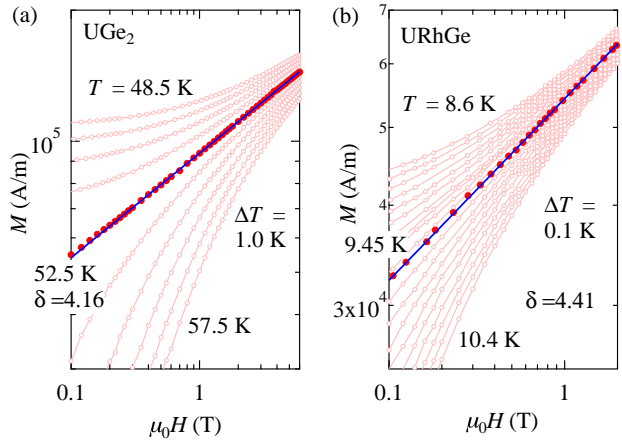}
\caption{\label{fig:epsart}(Color online) Magnetic field dependence of magnetization in (a)UGe$_2$ from 48.5 K to 57.5 K and in (b) URhGe from 8.6 K to 10.4 K. Dotted points indicate the critical isotherm at 52.5 K and 9.45 K for UGe$_2$ and URhGe, respectively. Blues lines show fits to eq. (3) to obtain the critical exponent $\delta$.}
\end{figure} 

Conventional methods to determine the critical exponents and the critical temperature involve the use of Arrott plots. Isotherms plotted in the form of $M^2$ vs. $H/M$ constitute a set of parallel straight lines around $T_{\rm C}$. The plot assumes that the critical exponents follow mean-field theory ($\beta$ = 0.5, $\gamma$ = 1.0, and $\delta$ = 3.0). The $H/M$ vs. $M^2$ plots in UGe$_2$ and URhGe do not yield straight lines around $T_{\rm C}$, indicating that the mean field model is not valid. So, we have re-analyzed the magnetization isotherms with the Arrott-Noakes equation of state which should hold in the asymptotic critical region\cite{arrott}.
   \begin{eqnarray}
  &&(H/M){^{1/{\gamma}}} = (T-{T_{\rm C}})/{T_1} + (M/{M_{1}})^{1/{\beta}}
   \end{eqnarray}
, where $T_1$ and $M_1$ are material constants. In the corresponding modified Arrott plot, the data for UGe$_2$ and URhGe are represented in the form of $M^{1/{\beta}}$ versus $(H/M)^{1/{\gamma}}$ as shown in Figure 1 (a) and (b).  The values $\beta$ and $\gamma$ are chosen in such a way that the isotherms yield as closely as possible a linear behavior.  A best fit of equation (4) to the data in UGe$_2$ for 47.0 K  $<$ $T$ $<$ 57.5 K and 0.1 T $<$ ${{\mu}_0}H$ $<$ 6 T yields $T_{\rm C}$ = 52.6 $\pm$ 0.1 K, $\beta$ = 0.334 $\pm$ 0.002, and ${\gamma}$ = 1.05 $\pm$ 0.05. The fit of the data in URhGe for 8.5 K  $<$ $T$ $<$ 10.4 K and 0.1 T $<$ ${{\mu}_0}H$ $<$ 2.0 T yields $T_{\rm C}$ = 9.44 $\pm$ 0.02 K, $\beta$ = 0.303 $\pm$ 0.003, and ${\gamma}$ = 1.02 $\pm$ 0.03. The obtained critical exponents are shown in Table. II.

   \begin{figure}[b]
\includegraphics[width=8.5cm]{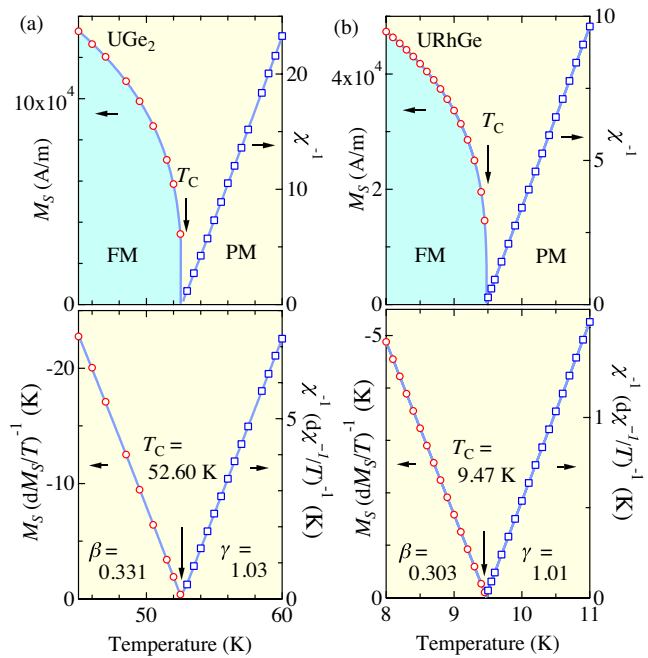}
\caption{\label{fig:epsart}(Color online) [Upper panels] Temperature dependence of the spontaneous magnetization $M{_s}(T)$ and the inverse of the initial magnetic susceptibility ${\chi}^{-1}$ determined from the modified Arrott plot and [Lower panels] Kouvel-Fisher plots for $M{_s}(T)$ and ${\chi}^{-1}$ in (a) UGe$_2$ and (b) URhGe.}
\end{figure} 
   The third critical exponent $\delta$ can be determined from the critical isotherm at $T_{\rm C}$ according to the eq (3) as shown in Figure 2 (a) and (b).  From fits to the isotherms at 52.5 K in UGe$_2$ and at 9.48 K in URhGe with eq. (3), the value of $\delta$ was obtained as $\delta$ = 4.16 $\pm$ 0.02 for UGe$_2$ and 4.41 $\pm$ 0.02 for URhGe. These values are lower than that expected for 3D Ising ferromagnet ($\delta$ = 4.80). The value of $\delta$ can be calculated from $\beta$ and $\gamma$ using Widom scaling relation ${\delta}$ = 1+${\gamma}/{\beta}$\cite{widom}. The value of ${\delta}$ was estimated as 4.15 $\pm$ 0.05 for UGe$_2$ and 4.37 $\pm$ 0.05 for URhGe. The values are consistent with those determined from the critical isotherms. 
   
   In UGe$_2$ and URhGe, the values of the critical exponent $\beta$ for the magnetization are close to those in the 3D ferromagnets. Meanwhile, the values of the exponents $\gamma$ for the magnetic susceptibility and $\delta$ for the critical isotherms are smaller than those expected for the 3D Ising model. 

\subsection{Kouvel-Fisher method}
The critical exponents $\beta$ and $\gamma$ can be more accurately determined by the Kouvel-Fisher (KF) relations\cite{kouvel}.
        \begin{figure}[t]
\includegraphics[width=8.5cm]{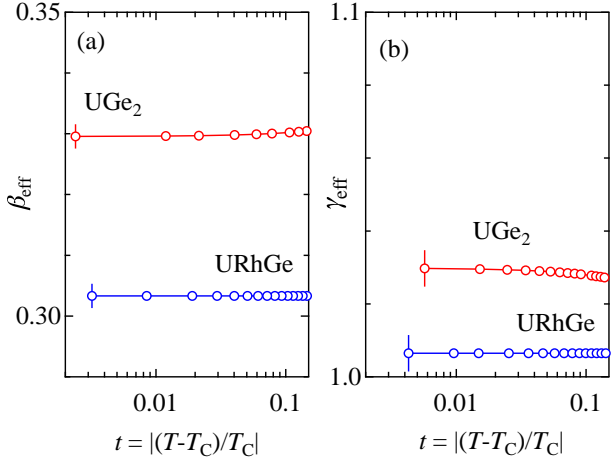}
\caption{\label{fig:epsart}(Color online)
Effective exponents (a) ${\beta}_{\rm eff}$ below $T_{\rm C}$ and (b) ${\gamma}_{\rm eff}$ above $T_{\rm C}$ as a function of reduced temperature $t$ [=$|({T}-{T_{\rm C}})/{T_{\rm C}}|$] in UGe$_2$ and URhGe.}
\end{figure}   

 In the modified Arrott plot, the straight lines intersect the $M^{1/{\beta}}$-axis in the ferromagnetic state at the values $M{_s}^{1/{\beta}}$ where $M{_s}$ is the spontaneous magnetization and in the paramagnetic state at ${\chi}^{-1/{\gamma}}$. The obtained temperature dependences of the spontaneous magnetization $M{_s}$ and the initial magnetic susceptibility ${\chi}$ are shown in upper panels of Figure 3 (a) and (b). Solid lines represent fits to the data with Eq. (2) and (1) for $M{_s}(T)$ and ${\chi}^{-1}(T)$, respectively. The KF method is based on following two equations:

       \begin{eqnarray}
 M{_S}(T)[dM{_S}(T)/dT]{^{-1}} &=& (T-T{_{\rm C}}^{-})/{\beta}(T)\\ 
  {{\chi}}{^{-1}}(T)[d {{\chi}}{^{-1}}(T)/dT]{^{-1}} &=& (T-T{_{\rm C}}^{+})/{\gamma}(T)
   \end{eqnarray}

 Eq. (5) and (6) are derived from Eq. (4) in the limit $H$ $\rightarrow$ 0 for $T$ $<$ and $>$ $T_{\rm C}$, respectively. The quantities ${\beta}(T)$ and ${\gamma}(T)$ are identical with the critical values $\beta$ and $\gamma$, respectively, in the limit $T$ $\rightarrow$ $T_{\rm C}$. According to the equations, the values of $\beta$ and $\gamma$ can be determined from the slope of $M{_s}(T)[dM{_S}(T)/dT]{^{-1}}$ and $ {{\chi}}{^{-1}}(T)[d {{\chi}}{^{-1}}(T)/dT]{^{-1}}$-plots, respectively, at $T_{\rm C}$ and the intersection with the $T$-axis yields $T_{\rm C}$ as shown in low panels of Figure 3 (a) and (b). Solid lines represent the fits to the data with Eq. (5) and (6). The exponents for UGe$_2$ are determined as $\beta$ = 0.331 $\pm$ 0.002 and $\gamma$ = 1.03 $\pm$ 0.02 with $T_{\rm C}$ = ($T{_{\rm C}}^{+}$ + $T{_{\rm C}}^{-}$)/2 = 52.60 $\pm$ 0.02 K by the KF method. The exponents for URhGe are determined as $\beta$ = 0.303 $\pm$ 0.002 and $\gamma$ = 1.01 $\pm$ 0.02 with $T_{\rm C}$ = 9.47 $\pm$ 0.01 K by the KF method. The critical exponents in the two compounds are consistent with those determined in the modified Arrott plot. The values of the critical exponent $\gamma$ in UGe$_2$ and URhGe are also close to unity in the KF method. 
       \begin{figure}[t]
\includegraphics[width=6.5cm]{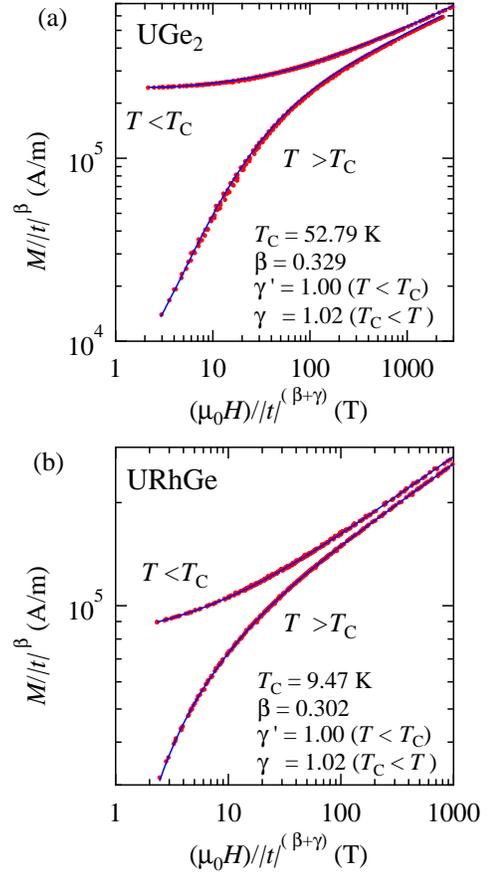}
\caption{\label{fig:epsart}(Color online)
Scaled magnetization as a function of renormalized field following Eq. (9) below and above the critical temperature $T_{\rm C}$ 
for (a) UGe$_2$ and for (b) URhGe. Solid lines show best fit polynomials. The magnetization data in the temperature range $t{\,}={\,}|({T}-{T_{\rm C}})/{T_{\rm C}}|{\,}< 0.1$ are shown.}
\end{figure}

 The critical exponents sometimes show various systematic trends or crossover phenomena as one approaches $T_{\rm C}$. This occurs if a magnetic system is governed by various competing couplings or disorders. To check this possibility, it is useful to obtain effective exponents ${\beta}_{\rm eff}$ and ${\gamma}_{\rm eff}$ as follows. 

    \begin{eqnarray}
{{\beta}_{\rm eff}} (t) =  d[{\rm ln}M{_s}(t)]/d({{\rm ln}{t}}),\\{{\gamma}_{\rm eff}} (t) =  d[{\rm ln}{{\chi}^{-1}}(t)]/d({{\rm ln}{t}})
   \end{eqnarray}
The effective exponents ${\beta}_{\rm eff}$ and ${\gamma}_{\rm eff}$ as a function of reduced temperature $t$ in UGe$_2$ and URhGe are plotted in Figure 4 (a) and (b). Both ${\beta}_{\rm eff}$ and ${\gamma}_{\rm eff}$ show a monotonic $t$-dependence in the asymptotic critical region, suggesting that the obtained exponents are not those that happen to appear around a crossover region between two universality classes as observed in Ni$_3$Al\cite{semwal}. 

\subsection{Scaling theory}
 We want to know whether the set of the critical exponents are the same below and above $T_{\rm C}$.  Analysis with scaling theory can determine separately the values ${\gamma}$' ($T<T{_{\rm C}}$) and $\gamma$ ($T{_{\rm C}}<T$). Theory predicts the existence of a reduced equation of state close to the ferromagnetic transition temperature\cite{privman}: 
    \begin{eqnarray}
  &&M({{\mu}_0}H, t) = {|t|^{\beta}} f{_{\pm}}({{\mu}_0}H/|t|^{{\beta}+{\gamma}})
   \end{eqnarray}
, where $f_{+}$ for $T{_{\rm C}} < T$ and $f_{-}$ for $T < T{_{\rm C}}$ are regular analytical functions. Defining the renormalized magnetization as $m$ $\equiv$ ${|t|^{-{\beta}}}{M({{\mu}_0}H, t)}$ and the renormalized field as $h$ $\equiv$ ${H}{|t|^{-({\beta}+{\gamma})}}$, the scaling equation is rewritten as $m$ = $f{_{\pm}}{(h)}$. This equation implies that $M({{\mu}_0}H, t)/{|t|^{\beta}}$ as a function of ${{\mu}_0}H/|t|^{{\beta}+{\gamma}}$ produces two universal curves: one for $T < T{_{\rm C}}$ and the other for $T > T{_{\rm C}}$ if the correct $\beta$, $\gamma$, and $t$ values are chosen. Figure 5 (a) and (b) show the scaled magnetization as a function of renormalized field following Eq. (9) at different temperatures below and above $T_{\rm C}$ in UGe$_2$ and URhGe, respectively. The magnetization data in the temperature range $t{\,}={\,}|({T}-{T_{\rm C}})/{T_{\rm C}}|{\,}< 0.1$ are shown. All data points fall on two curves in the two compounds. The scaling analysis yields $T_{\rm C}$ = 52.79 $\pm$ 0.02 K, $\beta$ = 0.329 $\pm$ 0.002, ${\gamma}'$ = 1.00 $\pm$ 0.02 for $T$ $<$ $T_{\rm C}$, and $\gamma$ = 1.02 $\pm$ 0.02 for $T_{\rm C}$ $<$ $T$ in UGe$_2$. The analysis yields $T_{\rm C}$ = 9.47 $\pm$ 0.01 K, $\beta$ = 0.302 $\pm$ 0.001, ${\gamma}'$ = 1.00 $\pm$ 0.01 for $T$ $<$ $T_{\rm C}$, and $\gamma$ = 1.02 $\pm$ 0.01 for $T_{\rm C}$ $<$ $T$ in URhGe.

   \begin{table*}[t]
\caption{\label{tab:table1}%
Critical exponents $\beta$, $\gamma$, and $\delta$, ferromagnetic transition temperature $T_{\rm C}$, and spontaneous magnetic moment ${\mu}_s$ in uranium ferromagnetic superconductors UGe$_2$, URhGe, UIr\cite{knafo,sakarya3}, and UCoGe\cite{huy,stock}. }
\begin{ruledtabular}
\begin{tabular}{cccccccc}
\textrm{}&
\textrm{$T_{\rm C}$ (K)}&
\textrm{$\beta$ }&
\textrm{${\gamma}'$ ($T<T{_{\rm C}}$)}&
\textrm{${\gamma}$ ($T{_{\rm C}}<T$)}&
\textrm{$\delta$}&
\textrm{${\mu}_s$ (${\mu}_{\rm B}$/U)}&\\
\colrule
UGe$_2$&&&&&&1.46&\\ 
Modified Arrott &52.6 $\pm$ 0.1  &0.334 $\pm$ 0.002 &  \multicolumn{2}{c}{1.05 $\pm$ 0.05} &&&\\ 
Kouvel-Fisher &52.60  $\pm$ 0.02  &0.331 $\pm$ 0.002 && 1.03 $\pm$ 0.02  &&&\\
Scaling &52.79  $\pm$ 0.02  &0.329 $\pm$ 0.002& 1.00 $\pm$ 0.02 & 1.02 $\pm$ 0.02 &&& \\
ln$(M)$ vs. ln${({{\mu}_0}H)}$ &  & & && 4.16 $\pm$ 0.02 && \\
\colrule
URhGe&&&&&&0.42\\
Modified Arrott &9.44 $\pm$ 0.02  &0.303 $\pm$ 0.002 &  \multicolumn{2}{c}{1.02 $\pm$ 0.03} \\
Kouvel-Fisher &9.47  $\pm$ 0.01  &0.303 $\pm$ 0.002 && 1.01 $\pm$ 0.02  \\
Scaling &9.47  $\pm$ 0.01  & 0.302 $\pm$ 0.001& 1.00 $\pm$ 0.01 & 1.02 $\pm$ 0.01  \\ 
ln$(M)$ vs. ln${({{\mu}_0}H)}$ &   & &&& 4.41 $\pm$ 0.02  \\
\colrule
UIr\cite{knafo,sakarya3} &45.15&0.355(50)   &  \multicolumn{2}{c}{1.07(10)} & 4.01(5)&0.5 \\
\colrule
UCoGe\cite{huy,stock} &2.5&   \multicolumn{4}{c}{$\sim${\,}Mean field type{\,}$\sim$ } & 0.05 \\
\end{tabular}
\end{ruledtabular}
 \end{table*}

 In previous neutron scattering experiments on UGe$_2$\cite{huxley1,kernavanois}, the value of $\beta$ was determined as 0.36(1) from the temperature dependence of the magnetic scattering intensity below $T_{\rm C}$ and mean field-like behavior of the magnetic susceptibility ($1/{\chi}{\,}{\propto}{\,}T$) was observed above $T_{\rm C}$. These are consistent with the present study. An important result of the scaling analysis is that set of the critical exponents in UGe$_2$ and URhGe are the same above and below $T_{\rm C}$. The values of $\gamma$ are close to unity below and above $T_{\rm C}$.

 The value of $\alpha$, the critical exponent for the specific heat ($C(T){\,} {\propto}{\,} |t|{^{\alpha}}$), is estimated as $\sim$ 0.3 for UGe$_2$ and URhGe using the Rushbrooke scaling relation (${\alpha}+2{\beta}+{\gamma}= 2$)\cite{rushbrooke}. This suggests the failure of the mean field theory ($\alpha$ = 0) where the specific heat does not show a divergent behavior at the transition temperature. In the theory, the contribution from the critical magnetic fluctuation to the specific heat becomes zero ($C_{mag}$ = 0) above $T_{\rm C}$. Meanwhile, the values of the magnetic specific heat $C_{mag}$ and the thermal expansion ${\alpha}_{mag}$ remain significant ($C_{mag}$ $>$ 0, ${\alpha}_{mag}$ $>$ 0) in the paramagnetic phase for the temperature range $t{\,}[=({T}-{T_{\rm C}})/{T_{\rm C}}]< 0.1{\,}{\sim}{\,} 0.2$ in UGe$_2$  (Fig. 4(a) and Fig. 7 in Ref. 32) and URhGe (Fig. 3 in Ref. 33). This suggests the development of critical fluctuations above $T_{\rm C}$. The experimental observations suggest that the mean field theory is insufficient to describe the thermodynamic quantities around $T_{\rm C}$. 
\section{DISCUSSION}

Table II summarizes the critical exponents $\beta$, $\gamma$, and $\delta$, ferromagnetic transition temperature $T_{\rm C}$, and spontaneous magnetic moment ${\mu}_s$ in uranium ferromagnetic superconductors. As mentioned in the introduction, the universality class for the 3D Ising model was expected from strong uniaxial anisotropy in the ferromagnetic magnetization of UGe$_2$ and URhGe. However, the present study suggests that the universality class in the compounds does not belong to any known class. While the values of $\beta$ are close to those in the 3D magnets, the values of $\gamma$ are close to unity, that expected from the mean field theory.

 The critical exponents in the ferromagnetic compound UIr with $T_{\rm C}$ of 46 K at ambient pressure are shown in Table II\cite{knafo,sakarya3}. Superconductivity has been found at high pressure in the ferromagnetic phase in UIr\cite{akazawa,kotegawa}. Although the magnetization shows strong uniaxial anisotropy in the ferromagnetic state in UIr,  the universality class of the critical phenomenon does not belong to the 3D Ising class\cite{knafo}. The values of the critical exponent $\beta$ and $\gamma$ are close to those in UGe$_2$ and URhGe. These results suggest a new universality class for the ferromagnetic transition in the uranium ferromagnetic superconductors.  In particular, the $T$-linear behavior of $\chi$ may be a characteristic feature. We discuss several possibilities for the unconventional critical scaling in these uranium ferromagnetic superconductors.

(1) A common feature in UGe$_2$, URhGe and UIr is that the crystal structure can be regarded as coupled chains of the uranium atoms (``zigzag structure") running along the $a$-axis in the orthorhombic structure of UGe$_2$ and URhGe, and along the $b$-axis in the monoclinic structure of UIr\cite{aoki2,akazawa}. The magnetic moments align parallel to the chain direction in UGe$_2$ and perpendicular to the direction in the latter two compounds. The nearest neighbor magnetic exchange interaction $J_{ij}$ = $J$ for bonds along the chain direction may differ from that ($J_{ij}$ = $rJ$ with $r$ $>$ 0) for bonds perpendicular to the direction. The magnetic structure can be mapped onto the anisotropic 3D Ising model. The critical exponents in the uranium ferromagnetic superconductors are not reproduced even when the spatial anisotropic exchange interaction is introduced\cite{yurishchev}.  Also, the present results are not consistent with numerical calculations on the anisotropic next nearest neighbor 3D Ising (ANNNI) model\cite{murtazev}.

 (2) Next, we discuss the itinerancy of the mobile $5f$ electrons. The long-range interactions of the delocalized magnetic moments yield the mean field theory-like behavior even very close to $T_{\rm C}$. It is necessary to know whether the analyses were done in the asymptotic critical region whose extent can be estimated by the Ginzburg criterion\cite{ginzburg2,chaikin,yelland}.
     \begin{eqnarray}
  &&{\Delta}{T_{\rm G}} = {T_{\rm C}}{k_{\rm B}^2}/[32{{\pi}^2}{({\Delta}C)^2}{{\xi_0}^6}]
   \end{eqnarray}

 Here, ${\Delta}C$ is the jump of the specific heat at the ferromagnetic transition temperature and ${\xi}_0$ is the magnetic correlation length. The Ginzburg criterion characterizes the temperature range where the mean field treatment does not hold. The stronger the itinerant character of the electron becomes, the narrower the asymptotic critical region. For example, the value of ${\Delta}{T_{\rm G}}$ in itinerant ferromagnet ZrZn$_2$ with the spontaneous magnetic moment  ${\mu}_s$ = 0.16 ${\mu}_{\rm B}$/f.u. and the magnetic correlation length ${\xi_0}$ = 33 {\AA} was estimated as ${\Delta}{T_{\rm G}}$ = 0.4 mK\cite{yelland,seeger}. The experimentally determined critical exponents in the temperature range $t{\,}={\,}|({T}-{T_{\rm C}})/{T_{\rm C}}|{\,}< 0.1$ are of the mean field type. The value of ${\Delta}{T_{\rm G}}$ in UGe$_2$ is estimated as $\sim$ 100 K using the neutron scattering and specific heat data\cite{huxley0,huxley1}. Therefore, our data are collected inside the asymptotic critical region where the mean field treatment fails. We cannot estimate ${\Delta}{T_{\rm G}}$ for URhGe and UIr since there has been no report on the correlation length ${\xi_0}$ in the two compounds. The analyses for URhGe and for UIr in Ref. 34 suggest that the temperature ranges of the asymptotic critical region $t{\,}[=|({T}-{T_{\rm C}})/{T_{\rm C}}|]$ are larger than 0.1 in the two compounds. The values of the exponent $\beta$ in UGe$_2$, URhGe and UIr are extremely smaller than that (${\beta}$ = 0.5) in the mean field theory. The observed $T$-linear behavior of 1/$\chi$ does not indicate that the analyses were done outside the asymptotic critical region. Furthermore,  the present analyses suggest no asymmetry in the temperature range of the asymptotic critical region in UGe$_2$ and URhGe. 

 In the ferromagnetic superconductor UCoGe, the spontaneous magnetic moment is 0.05 ${\mu}_{\rm B}$/U, one order of magnitude smaller than those in UGe$_2$ and URhGe\cite{saxena,aoki,huy}. The magnetic correlation length is estimated as ${\xi}{^{a,b}_0}$ = 86 {\AA} along the $a$ and $b$ directions and ${\xi}{^{c}_0}$ = 32 {\AA} along the $c$ direction\cite{stock}. The stronger itinerant character of the $5f$ electrons suggests a narrower asymptotic critical region in UCoGe. Indeed, the value of ${\Delta}{T_{\rm G}}$ is estimated as less than ${\Delta}{T_{\rm G}}{\,}{\sim}{\,}$1 mK from the specific heat and the neutron scattering data\cite{aoki2,stock}. The experimentally observed critical phenomenon is expected to be of mean field type. The $H/M$ vs. $M^2$ (Arrott-) plots yield straight lines around $T_{\rm C}$\cite{huy,ohta}. The critical behavior is masked by the strong itinerant character of the $5f$ electrons in UCoGe. 
 
(3) Even though a localized moment system, the universality class of the magnetic phase transition depends on the range of the exchange interaction $J(r)$. It is noted that the critical exponents for 3D Heisenberg, XY, Ising models and 2D Ising models in Table I are of short-range type, i.e., the magnetic interaction falls off rapidly with distance. Fisher {\it et al.} performed a renormalization group theory analysis of systems with the exchange interaction of a form $J(r)$ $\sim$ $1/r^{d+{\delta}}$, where $d$ is the dimension of the system and $\delta$ is the range of exchange interaction\cite{fisher}. Calculations showed that such a model for long-range interactions can hold for $\delta$ $<$ 2. The exponent $\gamma$ is given as $\gamma$ = ${\Gamma}$ ${\{}\delta$,$d$,$n{\}}$, where ${\Gamma}$ is a known function [Eq. (9) of Ref. 45] and $n$ is the dimension of the order parameter. This theory has been examined for different sets of ${\{}d:n{\}}$ ($d$, $n$ = 1, 2, 3), following a procedure similar to Ref. 46 which reported the critical phenomenon in Pr$_{0.5}$Sr$_{0.5}$MnO$_3$\cite{pramanik}. We do not find a reasonable solution of $\delta$ that reproduces the critical exponents in UGe$_2$, URhGe and UIr. 

(4) Classical dipole-dipole interaction affects the critical phenomenon. The case in gadolinium ($T_{\rm C}$ = 292.7 K, ${\mu}_s$ = 7.12 ${\mu}_{\rm B}$/Gd) has been extensively studied\cite{srinath}.  The effect of the dipole-dipole interaction in UGe$_2$, URhGe and UIr may be small since the strength of the effect is proportional to the square of the spontaneous magnetic moment ${\mu}_s$\cite{fisher2}. The critical exponents in the uranium ferromagnetic superconductors are not consistent with those of critical phenomenon associated with the isotropic or anisotropic dipole-dipole interaction\cite{frey1,frey2}.

(5) In metallic ferromagnets, the mean square amplitude of the local spin density $S_L^2$ persists in the paramagnetic state above $T_{\rm C}$ and  temperature change in its amplitude is the origin of the Curie-Wiess behavior in the magnetic susceptibility\cite{moriya}. The SCR theory gives ${\chi}^{-1}(T){\,}{\propto}{\,}(T-{T_{\rm C}}){^2}$ and $M{_S}^{2}(T){\,}{\propto}{\,}({T_{\rm C}^{4/3}}-T^{4/3})$ around $T_{\rm C}$, {\it i.e}, $\gamma$ = 2, $\beta$ = 1/2, and $\delta$ = 3, which are not consistent with those in UGe$_2$, URhGe and UIr. Some weak ferromagnets of the $3d$ transition metal such as MnSi or Co$_2$CrGa show an anomalous critical isotherm (${{\mu}_0}H{\,}{\propto}{\,}M^{1/{\delta}}$) with a value of $\delta$ close to 5\cite{bloch,nishihara}. The value is larger than that ($\delta$ = 3.0) in the SCR or the mean field theory. The behavior has been explained with the spin fluctuation theory by taking into account the zero point fluctuation under requirements of the total spin amplitude conservation (TAC) and the global consistency (GC)\cite{takahashi1,takahashi2}. Although the values of $\delta$ in UGe$_2$, URhGe and UIr are larger than that in the mean field theory, the $T$-linear behavior of 1/$\chi$ is not consistent with the theory ($\gamma$ = 2). We hope that the uniaxial magnetic anisotropy will be taken into account in the spin fluctuation theory. 
   
  As discussed in (1)-(5), the anomalous critical exponents in UGe$_2$, URhGe and UIr cannot be explained with previous approaches to critical phenomenon. The present study suggests that the ferromagnetic correlation between the $5f$ electrons differs from that in the 3D Ising system and this difference is a key point for the understanding of the ferromagnetism where superconductivity emerges.  Finally, we propose following two viewpoints on the ferromagnetism in the uranium superconductors for future studies on the anomalous critical scaling.

(i) We mention the dual nature of the $5f$ electrons between itinerant and localized characters in UGe$_2$\cite{yaouanc,sakarya,troc}.  The non-Landau damping for the magnetization density $\Gamma$ in UGe$_2$ and UCoGe has been discussed theoretically based on the duality\cite{mineev,chubukov}. Theoretical models for the superconductivity in the ferromagnetic state in URhGe and the antiferromagnetic state in UPd$_2$Al$_3$ have been developed based on the duality model\cite{sato,thalmeier,hattori2}. The duality may be a key point for the co-existence of the ferromagnetism and the superconductivity. The correlation length of the itinerant component with a magnetic moment of $\sim$ 0.02 ${\mu}_{\rm B}$/U was estimated as $\xi$ = 84 {\AA} in UGe$_2$ by the Muon spin rotation spectroscopy. This value is significantly larger than that found in the neutron scattering experiment whose main contribution comes from the localized component since the magnetic scattering intensity is proportional to the square of the magnetic moment\cite{huxley1}. A novel type of critical phenomenon may appear due to the two correlation lengths as well as a Hund-like coupling between the two components. 

(ii) UGe$_2$ has a tricritical point where the paramagnet to ferromagnet transition changes from a second-order to a first order phase transition when driven toward the ferromagnetic QCP by applying external pressure\cite{taufour}, similar to several itinerant ferromagnets such as ZrZn$_2$\cite{uhlarz}, Co(S$_{1-x}$Se$_x$)$_2$\cite{goto}, and MnSi\cite{thessieu}. This change of the nature of the transition has been regarded as a phenomenon specific to the quantum phase transition\cite{belitz}. Recently, the pressure effect on the ferromagnetic transition has been re-considered from different points of view. When the ferromagnetic transition temperature is strongly pressure-dependent, the magneto-elastic interaction or the critical fluctuation of the order parameter provides development of the first order instability at the phase transition\cite{mineev2,gehring,shopova}. Neutron Larmor diffraction study reveals that magneto-elastic coupling is strengthened at the pressures where superconductivity appears in UGe$_2$\cite{sokolov}. Mineev shows that the order parameter fluctuations give rise to the logarithmic increase of the specific heat near $T_{\rm C}$ in the uranium ferromagnetic superconductors\cite{mineev2}. As mentioned before, the magnetic specific heat $C_{mag}$ and the magnetic thermal expansion ${\alpha}_{mag}$ in UGe$_2$ and URhGe show the anomalous temperature dependence just above $T_{\rm C}$\cite{hardy,sakarya2}. Future theoretical study is necessary to determine for the effect of the order parameter fluctuations on the magnetization around $T_{\rm C}$.

 \section{conclusion}
 A dc magnetization study has been done of the critical phenomenon around the ferromagnetic transition temperature in high-quality single crystals of uranium ferromagnetic superconductor UGe$_2$ and URhGe. We have determined the critical exponents, $\beta$ for the magnetization, $\gamma$ for the magnetic susceptibility, and $\delta$ for the magnetic isotherm at the transition temperature with a modified Arrott plot, a Kouvel-Fisher plot, and the scaling analysis. Although the magnetization shows strong uniaxial magnetic anisotropy, the universality class of the critical phenomenon does not belong the three dimensional (3D) Ising system.  In the asymptotic critical region, the values of $\beta$ in UGe$_2$ and URhGe are close to those in 3D magnets but the susceptibility $\chi$ shows a mean field-like behavior (1/$\chi$ ${\propto}$ $T$). Similar critical exponents have been reported previously in 3D Ising ferromagnet UIr where the superconductivity appears under high pressure. The critical behavior may be limited to a very narrow Ginzburg critical region of ${\Delta}{T_{\rm G}}{\,}{\sim}$ 1 mK because of the strong itinerant character of the $5f$ electrons in the ferromagnetic superconductor UCoGe. The anomalous critical exponents in UGe$_2$, URhGe and UIr cannot be explained via previous approaches to the critical phenomena. We suggest that this unconventional critical scaling of magnetization is inherent in the uranium ferromagnetic superconductors and it reflects a peculiar feature of the ferromagnetism of the $5f$ electrons where superconductivity emerges.

   \section{ACKNOWLEDGMENTS}
We acknowledge discussions with Drs. H. Sakai, K. Kubo, N. Metoki, Y. Tokunaga and K. Kaneko. We also thank Dr. S. Kambe and Prof. V. P. Mineev for giving suggestive comments which improve this paper. This work was supported by a Grant-in-Aid for Scientific Research on Innovative Areas ``Heavy Electrons (Nos. 20102002 and 23102726), Scientific Research S (No. 20224015), A(No. 23246174), C (Nos. 21540373, 22540378 and 25400386), and Young Scientists (B) (No. 22740241) from the Ministry of Education, Culture, Sports, Science and Technology (MEXT) and Japan Society of the Promotion of Science (JSPS).

\bibliography{apssamp}

\end{document}